# Big Data Analytics-Enhanced Cloud Computing: Challenges, Architectural Elements, and Future Directions

Rajkumar Buyya[1], Kotagiri Ramamohanarao[1], Chris Leckie[1], Rodrigo N. Calheiros[1], Amir Vahid Dastjerdi[1], and Steve Versteeg[2]

[1]Cloud Computing and Distributed Systems (CLOUDS) Laboratory
Department of Computing and Information Systems
The University of Melbourne, Australia
E-mail: {rbuyya, kotagiri, caleckie, rnc, amir.vahid}@unimelb.edu.au

[2]CA Technologies, Melbourne, Australia
steve.versteeg@ca.com

*Abstract*—The emergence of cloud computing has made dynamic provisioning of elastic capacity to applications on-demand. Cloud data centers contain thousands of physical servers hosting orders of magnitude more virtual machines that can be allocated on demand to users in a pay-as-you-go model. However, not all systems are able to scale up by just adding more virtual machines. Therefore, it is essential, even for scalable systems, to project workloads in advance rather than using a purely reactive approach. Given the scale of modern cloud infrastructures generating real time monitoring information, along with all the information generated by operating systems and applications, this data poses the issues of volume, velocity, and variety that are addressed by Big Data approaches. In this paper, we investigate how utilization of Big Data analytics helps in enhancing the operation of cloud computing environments. We discuss diverse applications of Big Data analytics in clouds, open issues for enhancing cloud operations via Big Data analytics, and architecture for anomaly detection and prevention in clouds along with future research directions.

## I. INTRODUCTION

Information Technology (IT) systems are now critical components of almost every business, government, and societal applications. IT does not just support the business, but IT itself is the business in many instances. Today there are few business services on various IT systems for their delivery. For example, banks rely on IT systems to process financial transactions, airlines rely on IT for their ticketing and timetabling, and even tollways rely on IT for their billing and safety systems. For most companies, when the IT systems are down, the company cannot do business. When the IT systems are not adequately designed, implemented or provisioned, the company can lose its business and customer satisfaction.

For IT to do its job in delivering business services, it needs to meet the following requirements:
- The IT systems need to be fortified against outages. Systems need to be provisioned with sufficient capacity to meet even the peak demand. If the capacity is insufficient business opportunities will be lost. If the capacity is too great, operating overheads are increased.
- The right services need to be provided to maximize business value.
- Systems need to be configured for maximum efficiency and robustness.

However, meeting these objectives in the modern enterprise environment is a challenging task. First, IT systems are increasingly complex and interdependent. To deliver a specific business service, there will be typically many different systems involved, each of which may be communicating and reliant on other third party systems. The complexity and interdependence are the causes of many possible failures, making it difficult to implement robust services and to diagnose the root cause of failures. Second, there is a high degree of uncertainty in workload demand. Demands are rarely uniform and predictable, but they tend to be highly irregular, bursty, and spiky in nature. In particular, external events and anomalies can cause radical shifts in service demand. As these events are very difficult to predict, the demand at a given point in the future will be unknown. System capacity is therefore likely to be over-provisioned or under-provisioned in the face of unexpected events.

The rise of cloud computing made dynamic provisioning of elastic capacity on-demand possible for applications hosted on data centres. This is because cloud data centers contain thousands of physical servers hosting orders of magnitude more virtual machines that are allocated on demand to users in a pay-as-you-go model.

Cloud data centers provide this elasticity through the notion of Infrastructure on Demand (IoD). However, not all systems are able to be scaled up by just adding more virtual machines. For these systems to be effective, they should be designed to be able to exploit IoD. Furthermore, even for systems that are able to exploit IoD, there is typically a delay in bringing new capacity online. Therefore, it is essential, even for scalable systems, to project workloads in advance rather than using a purely reactive approach.

From the above discussion, it can be noticed that current systems can benefit from the capacity of *prediction* of future application demand, *infer* the effect of such demand in the infrastructure and, consequently, in the applications, and *detect anomalies* in the infrastructure and applications in real time. These are problems addressed by researchers and practitioners in the areas of *machine learning* and *data mining*. Nevertheless, given the nature of modern cloud infrastructures (thousands of physical servers and virtual machines) generating real time monitoring information, along with all the information generated by operating systems (logs, system calls, etc.), applications (response times, latency), and user behavior (click analytics), this data reaches volume, velocity, and variety that is not efficiently handled by traditional machine learning and data mining techniques.

By leveraging Big Data techniques and technologies (large-scale data mining, time-series analysis, and pattern mining), data such as event and log data can be captured at finer granularity with longer histories and analyzed in multiple projections. In this paper, we propose how the application of Big Data Analytics can enhance the operation of cloud computing infrastructures. We present various applications of Big Data analytics in clouds, open issues for enhancing cloud operations via Big Data analytics, an architecture able to tackle the problem of anomaly detection and prediction in clouds, and future research directions.

The rest of this paper is organized as follows. Section II presents the motivation for this work. Sections III, IV, and IV discuss, respectively, the problems of anomaly detection and prevention, workload and performance prediction, and clustering, all in the context of improving operations of cloud computing services. Section VI presents architecture for anomaly detection and prevention in clouds, which is evaluated in Section VII. Section VIII presents future research directions in the topic and Section VIX concludes the paper.

II. MOTIVATION

Cloud computing enables users to acquire computational resources as services in a pay-per-use model, and this is generally called Infrastructure on Demand (IoD). The exact IoD that is commercialized as a service varies in one of three service models: Infrastructure as a Service (IaaS), Platform as a Service (PaaS), and Software as Service (SaaS). Each of these models provides a different view for users of what type of resource is available and how it can be accessed.

In the IaaS model, users acquire virtual machines that run in the hardware of cloud data centers. Virtual Machines (VMs) can contain any operating system and software required by users, and typically users are able to customize the VMs to their own needs. Typically, IaaS providers charge users by the time that VMs run, and the exact cost per unit of time depends on the hardware resources (memory, CPU cores, CPU speed) allocated to the VM, which users can select among different amounts offered by providers. Therefore, the views users have of the system are restricted to Operating System and above levels.

In the PaaS model, users are provided with an environment where applications can be deployed. At this level, users are able to collect metrics about application-level resource usage and performance. At the SaaS level, users access an application, being usually charged on a subscription basis. Metrics available at this level (if any) regard application-specific data.

The view that cloud providers have, often, are those one level below the view that users have: IaaS cloud providers have metrics available of the platform level (e.g., resource usage of physical hosts), PaaS providers have infrastructure-level information (e.g., container-level resource usage) and SaaS providers have platform-level information (e.g., response time of requests to the application).

This different views and objectives of analysis of data affect the techniques that can be applied, their scope, and their results, as we discuss in the next sections.

III. ANOMALY DETECTION AND PREVENTION

Service Level Agreements (SLAs) are one important aspect of the engagement between cloud service providers and cloud users. Because there is a strong competition among providers in all service models, the damage to the reputation of a provider resulted from violating SLA terms can be substantial: It not only leads to penalties applied to providers, but also the risk of having the news of bad user experience spread through social media, resulting in loss of potential customers (and even existing ones).

Therefore cloud service providers need to strive to meet SLAs. However, given the complexity and scale of cloud infrastructures, it is challenging for providers to guarantee that all systems and software are working according to the desired specification at all times. Thus, it is important that providers have mechanisms in place to detect abnormal activity in their infrastructure, platform, and software.

In the context of this paper, we use the term "anomaly detection" to refer to detection of patterns of utilization of resources and metrics that deviate from the expected value. This can be caused by failures in hardware and software components, but also by an excess of users of the applications at higher levels.

A challenging aspect of anomaly detection in clouds concerns the fact that effective methods need to be unsupervised [1]. This is because the variety of hardware, services, and applications, along with variation in application demand generate much more data that can be labelled by experts.

Regarding anomaly detection for IaaS, earlier efforts in this direction [2][3] are suitable for small scale private clouds, but are not scalable enough to support state-of-the-art large-scale data centers that have many orders of magnitude more resources to be managed. This is because these early approaches assume communication models, such as all-to-all that are not scalable, or use methods such as k-Nearest Neighbors (k-NN), which have high asymptotic complexity and thus cannot generate output in the speed required for proper SLA compliance.

Other approaches for the problem are based on time-series analysis of the data [1]. These approaches operate with the assumption that patterns that are time-dependent emerge in the utilization of cloud services. Therefore, the time dimension cannot be ignored in the anomaly detection process. These approaches are achieving a degree of success in identifying anomalies in a single variable (usually, CPU). However, it is desirable that multiple attributes (i.e., memory, storage, network along with CPU) are considered at the same time to reduce the number of false-positives.

A related problem to anomaly detection is anomaly prevention, which requires underlying support systems to detect and respond to the anomaly in the earliest possible time. This is challenging due to the sheer volume of monitoring data generated by large-scale data centers require near real-time solutions. PREPARE [4] achieves that for private IaaS. However, the suitability of the approach for large-scale public clouds is yet to be investigated.

At a higher layer, PerfCompass [5] has been developed with the goal of detecting performance anomalies in applications. It tracks frequency and runtime of system calls to detect abnormal behavior of applications and to estimate the source of the fault as being internal or external to the VM hosting the offending service.

In common with all approaches, there is the need to be able to define what an anomaly is and, in case of unsupervised learning, finding mechanisms that increase the accuracy of the method and provide a timely output.

***Open Problem #1:*** *Dealing with unseen anomalies.* Most approaches for anomaly detection and prevention in the cloud build the anomaly detection models based on the probability distribution of previous state information. Research is required on a reactive method that can dynamically decide when models need to be updated or built. Reactive approaches particularly contribute to the anomaly detection in real time, especially for the cold start period. For example, Self-Organizing Maps can be used in early stages, as it is capable of capturing complex system behavior while being computationally less expensive than comparable approaches such as k-nearest neighbor. In addition, sentiment analysis on Web and social networks data can be used to correlate the system anomalies with the behavior of web applications' users in the cold start period, helping to differentiate anomalies caused by hardware issues from anomalies caused by used behavior.

## IV. WORKLOAD AND PERFORMANCE PREDICTION

Techniques for prediction can be used in the context of cloud computing to help providers to optimize the utilization of resources. The rationale behind the idea is that, by correctly estimating the future demand for resources (by correctly predicting the expected workload of an application or service), the right amount of resources that delivery the expected SLA with minimum wastage in resource utilization can be achieved.

These techniques, follow *proactive* approach, contrast with *reactive* approaches used in the management of cloud resources. Reactive approaches apply actions to decide the right amount of resources *after* issues with performance are detected. These techniques usually apply anomaly detection techniques discussed in the previous section.

Neves et al. [6] developed Pythia, a system that predicts the communication needs of a MapReduce application, and then reconfigures the network, to optimize bandwidth allocation to the application.

Islam et al. [7] applies neural network and linear regression to predict the moment where more resources will be necessary, what leads to the need of new virtual machines will be required. Thus, the boot process of VMs can be initiated before the need for resources, reducing the risk of SLA violations.

Davis et al. [8] investigated the use of linear regression to predict resource utilization in clouds. Authors found that a weighted multivariate linear regression presented (MVLR) low average errors for short-term prediction with trends in the time series. Seasonality could be handled with an ensemble of scaled Fourier transform, MVLR, and weighted regression.

Even in the case that system can correctly estimate how the load of applications and services will expand or shrink, there is still the need to apply corrective actions. This in turn requires systems to be able to estimate how changes in the underlying infrastructure (e.g., number of machines dedicated to an application and amount of CPU, memory, and network of such machines) will affect the performance of the elements in the upper layers.

In common among all the approaches, there is the assumption that the states of infrastructure-level resources (e.g., CPU, memory, disk, and network) are good predictors of the state of the applications in the upper layers. This raises the first open problem we identify in this area.

***Open Problem #2:*** *Correlation between infrastructure performance and application performance.* Although intuitively one might expect a strong correlation between resource utilization at the infrastructure level and performance of applications, this assumption is based on too many simplifications. Firstly, it fails to consider the true nature of applications: complex cloud applications can experience stages of intense CPU activities and intense communication activity. This can be hard to be captured by single variable approaches. Also, the excess of load can be caused by some perturbation in the application itself (e.g., flash crowd) that cannot be solved solely by infrastructure-level actions, but may require more complex actions at the platform level (e.g., scaling out and load balancing). Correlation of data of multiple levels to provide a holistic view of the system with the goal of improving performance of applications and meeting SLAs is an open question that needs to be investigated in more details.

## V. CLUSTERING

Clustering is an unsupervised learning technique that enables grouping of objects by similarity: objects sitting in the same cluster are more similar among themselves than objects in different clusters [9].

Clustering has been applied in the context of cloud computing to enable optimization of execution of tasks (i.e., requests for the execution of applications, usually batch applications). In particular, clustering of tasks and jobs (i.e., group of tasks that are handled as a single unit) obtained from traces generated by Google have been used for identification of similarities regarding resources requirement and execution time [10][11][12]. This helps in the selection of machines where tasks should be executed and to estimate the execution time, an activity that is required to enable optimal scheduling of tasks in the available resources.

Clustering has also been used to help in the problem of placement of large amounts of data required by scientific applications hosted in the cloud [13]. The aim of the work is reducing the amount of data movement required by the application, which also helps in reducing execution time of applications. Indirectly, it also helps in reducing network usage of data centers, what also contributes to improve overall performance of applications hosted in the data center.

In a similar way, clustering has been also shown to be successful in helping in the problem of live migration of virtual machines [14]. Live migration consists in transferring a running virtual machine from one physical server to another while it is running (i.e., without perceivable interruption on the services provided by the applications running on the VM) [15]. In this particular approach, the goal was inter-cloud live migration, which means that the source and destination servers where located in different data centers. Clustering has been used to determine which machines should be simultaneously migrated.

***Open Problem #3:*** *Other applications of clustering for resource management in cloud data centers.* So far, clustering has been much less explored than prediction and regression. Given the scale of cloud infrastructures, clustering may be a valuable tool in reducing the complexity of management, by helping actions to be taken on a cluster-basis rather than on more fine-grained basis. Therefore, if meaningful ways of classifying resources can be found, that are more coarse-grained than per user or per resource type (things that are known *a priori* and therefore do not require application of clustering), it is likely more efficient resource management might be achieved.

## VI. AN ARCHITECTURE FOR ANOMALY DETECTION AND REACTION IN IAAS CLOUDS

Decisions towards selecting the appropriate cloud provider, the type of resource for a given application, number of cloud resources, and the moment when such resources should be requested have to be made by the user. Together, these activities are referred to as "dynamic cloud provisioning". There is a lack of research/advances made in provisioning driven by prediction, detection, and reaction to anomalies. This is due to the system administrator's inability to scale the system if an abnormal peak of demand occurred before the development of cloud computing [16][17]. As cloud computing enables the infrastructure to be dynamically scaled, a new opportunity for achieving high Quality of Service (QoS) emerged. At the same time, as utilization of cloud resources incurs financial cost, scaling of resources should be the minimum possible that satisfies the business needs that is highly dynamic and unpredictable.

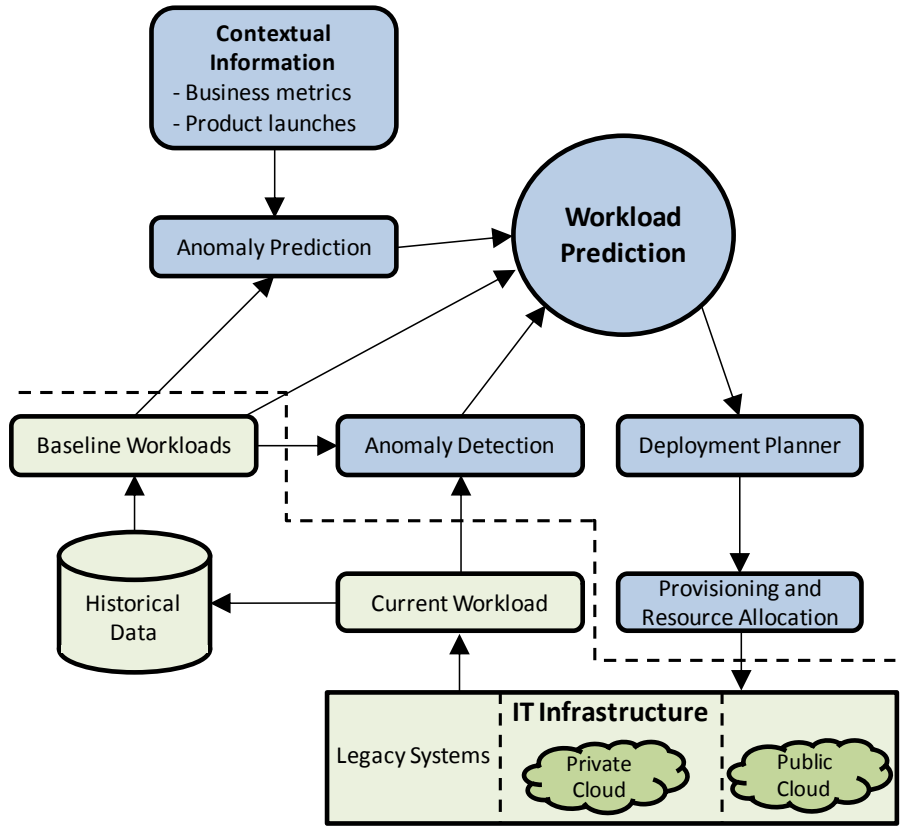

Fig. 1. Architecture for anomaly detection and reaction in clouds.

The high level architecture of the framework is shown in Figure 1. The dashed lines limit the interface of the architecture, which is composed of Anomaly Prediction, Anomaly Detection, Workload Prediction, Deployment Planner, Provisioning and Resource Allocation, and Contextual Information. Besides these core elements of the architecture, the following are sources of information and external systems that support such architecture:

- *Contextual information*: data used by our proposed system architecture to make decisions are supplied from different sources, such as: logs from the infrastructure operations (which may indicate unexpected behavior in the system); information about release of new products (that may cause an extra load of consumers interested in learning or buying such products); business metrics related to expected performance parameters of the system; and data from social networks (that may indicate the sentiment of customers to a new product and may affect the input workload of the system).
- *Baseline workloads*: the baseline workloads are built with the patterns observed from historical data, and enable the determinations of fluctuations in the system input along the time. Such workloads provide insights on how the demand changes according to the period of the day, day of the week, season, months, etc.
- *Current Workload*: this is the observed workload in the system in a given moment and it is acquired via monitoring tools. This information is constantly logged as historical data for future use. The framework uses this log to emulate real time loads to our proposed framework in order to enable the detection of ongoing anomalies.
- *IT Infrastructure*: the target IT infrastructure for our proposed framework consists of a hybrid cloud, composed of both public cloud providers and in-house infrastructure owned by the cloud service provider as well as legacy systems (either hardware or software) that are not cloud-ready.

*A. Anomaly Prediction*

The anomaly prediction module is responsible for estimating a possible anomaly in the workload to be received by the cloud service provider in a future moment and the confidence level in the occurrence of such event. It has two main sources of information: the baseline workloads, which supply an estimation of a typical workload for a particular time, day, and period of the year; and contextual information. This log data is used to build an analytical model using Markov models. The Markov model can be parametrically changed to create anomalous behaviors to study the robustness of our proposed system.

Different sources of data may have different degrees of structure, and they can be available in different formats. Modelling them as Markov models overcome compatibility issues such as data formats and dimensionality. Therefore, the following actions are essential for enabling a correct and timely operation of the Anomaly Prediction module:

- Selection of appropriate sources of data for prediction;
- Filtering of data, so only data from relevant sources are considered for the prediction and modelling;
- Extraction of data of interest from the filtered data, including Big Data analysis and data mining;
- Actual prediction of the expected workload;
- Actual prediction of failures in the system;
- Determination of prediction confidence levels.

Another important aspect to be considered during the anomaly prediction is that the result of the prediction must be timely, so that there is enough time for the rest of the components of the system to react.

*B. Anomaly Detection*

Because predictions are not always accurate, and unpredictable circumstances may affect the workload beyond a level that can be predicted, a second line of defense against loss of performance caused by anomalous workloads or failures in the system needs to be considered.

In our framework, this second line of defense is carried out by the Anomaly Detection module. Operation of this module is based on the workload observed in a given time and baseline workloads. When these two measurements diverge by a specific margin, an alarm is triggered by this module to the Workload Prediction module.

This is achieved with anomaly detection algorithms that analyze the described data to make a decision about the severity of the anomaly and the likelihood of its transiency. This is important because, if the anomaly is expected to incur for a short period of time, it is possible that it ceases before the environment finishes its scaling process to handle it. Furthermore, if the anomaly is not severe, it is possible that the available resources are able to handle it without the need of more resources. In this case, no alarm should be triggered and the system should keep its current state.

*C. Workload Prediction*

The earlier modules (Anomaly Detection and Anomaly Prediction modules) focus in determining patterns that may lead to an increased (or decreased) interest of users to applications hosted by the cloud service provider, an estimation of such interest, and the risk of failures in the system leading to anomalous behavior of the systems. It does not directly translate to a quantifiable measurement of performance of the system because of the unexpected workloads.

The Workload Prediction module carries out the translation of observed or unexpected variance in estimations to the business impact of possible disruptions. To achieve this, this module quantifies the expected workload in terms of requests per second along a future time window and combines this information with business impacts. Therefore, the output generated by this module (and the algorithms to be developed as part of its conception) is concrete business metrics that have value to managers of IT infrastructures.

*D. Deployment Planning*

The Deployment Planning component of our framework is responsible for advising actionable steps related to deployment of resources in a cloud infrastructure to react to failures or anomalies in the system. Automation engine in the Provisioning and Resource Allocation module of the system executes these steps.

The tasks performed by this module are challenging as the goal of such plan is to mitigate the effect of variations in the system that disturb its correct operation. Correcting such anomalies means re-establishing a QoS level to users of the affected platform. However, enabling QoS requirements driven execution of cloud workloads during the provisioning of resources is a challenging task. This is because there is a period of waiting time between the moment resources are requested and the provision of resources by the cloud providers and the time they are actually available for workload execution. This waiting time varies according to specific providers, number of requested resources, and load on the cloud.

As our framework cannot control waiting times, this time has to be compensated by other means. Possible approaches are increasing the number of provisioned resources to speed up the workload delayed because of delays in the provisioning process or to predict earlier the resource demand albeit with low accuracy and probability. However, the first solution may not resolve the problem for most web applications because users affected by the delays are likely to abandon the access to the service, which results in loss of opportunity for revenue generation in the affected system. Another challenge for the deployment planning process concerns selection of the appropriate type of resource to be allocated. Our second approach overcomes the problem but may be slightly more expensive due to potential overprovisioning of resources. As our proposed algorithms are based on learning techniques, these methods are likely to improve their quality over the time by observing the performance of the system.

Different cloud providers have different offers in terms of combination of CPU power, number of cores, amount of RAM memory, and storage of their virtual machines. Providers also offer multiple data centers in different geographic locations. This affects the expected latency for communication and data transfer between users and resource and consequentially observed response times. Therefore, the Deployment Planning module needs to describe resource in a vendor-agnostic way, so the Provisioning and Resource Allocation module can translate the description to a concrete vendor offer once a vendor is selected.

*E. Provisioning and Resource Allocation*

This module is responsible for the enactment of the provisioning planning generated by the framework. It interacts with different public cloud providers and resource management system of the public cloud in order to enable the realization of the planning decision performed by the Deployment Planner module. Furthermore, different combinations of features have different costs. In order to meet user budget constraints, the planning algorithm has to take into account the combination of resources that meet performance requirement of the estimated workload at the minimum cost. More specifically, this component has the following functions:

- Translation of resource requirements from a vendor-agnostic description to specific offers from existing cloud providers;
- Selection of the most suitable source(s) of resources based on price, latency, resource availability time, and SLA;
- If possible, perform automatic negotiation for better offers from providers with compromising SLA.

***Open Problem #4:*** *Leveraging existing Big Data ecosystem to implement advanced analytics solutions supporting Big Data-enhanced cloud computing.* There is huge ecosystem of (sometimes competing) Open Source technologies that are widely adopted for all layers of Big Data analytics. This include Hadoop/YARM (MapReduce and other parallel programming models), Storm (stream processing), Spark (analytics), Pig and Hive (high level query languages), Mahout (high level analytics tasks), and Cassandra, CouchDB, BlinkDB, HDFS (file systems and NOSQL databases). The question is how to leverage these tools to enable complex analytics that enables SLAs to be met with minimum resource consumption.

VII. PERFORMANCE EVALUATION

In this section, we present an evaluation of the conceptual framework described in the previous section. In particular, we focus in the *Workload Prediction* module, which is the core of the proposed framework.

This evaluation experiment leverages the methodology we utilized in our earlier work [18]. However, it focusses on *different objectives,* and utilizes a *different dataset* from the same source.

The workload utilized has been obtained from the page view statistics from all Wikimedia projects [1] on 1$^{st}$ of September, 2014. The information is organized by the language of the accessed document (web page, figure, text file, etc.). We use the data about http access to Chinese language documents. To gain insight of the traffic for each project for the whole day we analyzed traces which consist of 24 compressed files each containing 160 million lines (around 8 GB in size). We utilized Map-Reduce on a cluster of 4 nodes to calculate the number of requests more effectively and faster.

The traces provide hourly access information and they were converted to access per second using a log-normal distribution, and then consolidated in 5-seconds intervals for processing purposes. We utilize the first 17 hours for training purposes and the next 8 hours for testing. Because of the time-series nature of the workload, we utilize the ARIMA method [19] for fitting and prediction. This method decomposes the time-series into three components. The first is an *Autoregressive* component of *order p* that models a point as a linear combination of $p$ previous observations. This component can be *Integrated d times* to eliminate stationarity (as ARIMA processes need to be non-stationary). The third component is a *Moving Average* of order $q$ component that models a point as a linear combination of the $q$ previous observation errors.

The fitting process is carried out with different *(p, q, d)* parameters. The module needs to use the values that minimize some error measurement (for example, *Mean Square Error*) to increase the framework's accuracy and these value are likely to be different for different workloads (which, in the context of this experiment, are the different languages of documents accessed by users). Errors are presented in Table I, whereas Figure 2(a) show the effect of different parameters in the fitting process for all the training

---

[1] Available at http://dumps.wikimedia.org/other/pagecounts-raw/

period. To enable a better visualization, Figure 2(b) presents a 1-hour snapshot of the same data.

Next, we evaluate the effect of the difference in accuracy among the models in the quality of the provisioning and consequently in the Quality of Service to end user. We utilized the CloudSim simulator [20] to model a data center with 500 computers, each with 8 cores, 16 GB of RAM and 1 TB of storage. In the simulation, workload has been submitted according to the traces and the provisioning has been carried out according to the prediction of each mode. Provisioning is carried out with virtual machines that use 1/8 of the available resources of the host, so each VM has exclusive access to one core. A routine for adjusting the provisioning is invoked every 5 minutes for a period of 5 minutes ahead, so there is enough time for new VMs to be started if necessary. Each request is assumed to take 100ms and the target QoS for response time is 500ms.

The provisioning is carried out based on the estimated load by the different ARIMA models, following the procedure introduced in our previous work [21]. The output metric are the number of VM hours require to process the workload (normalized by the number of hours required to meet the QoS via static provisioning). Furthermore, because prediction techniques are subject to inaccuracies, we also report the normalized number of requests rejected due to prediction errors.

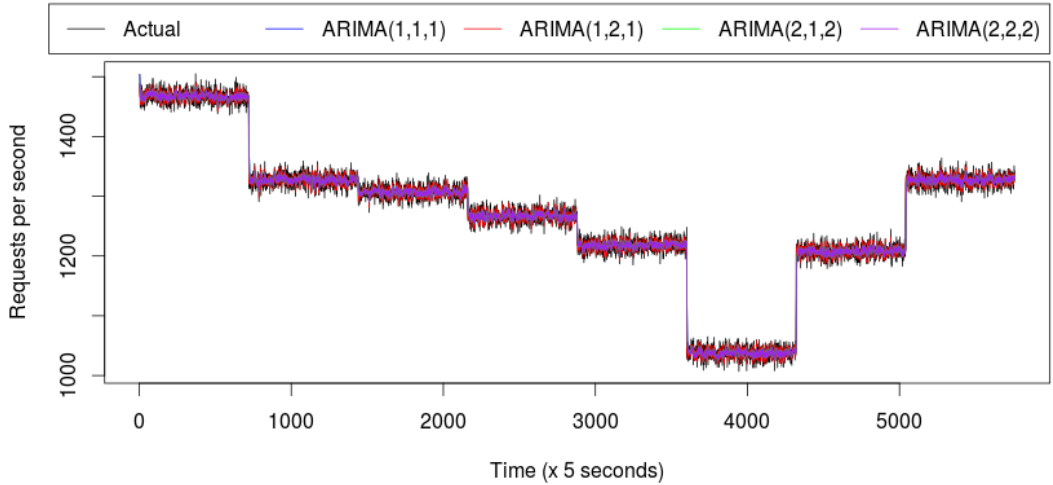

(a)

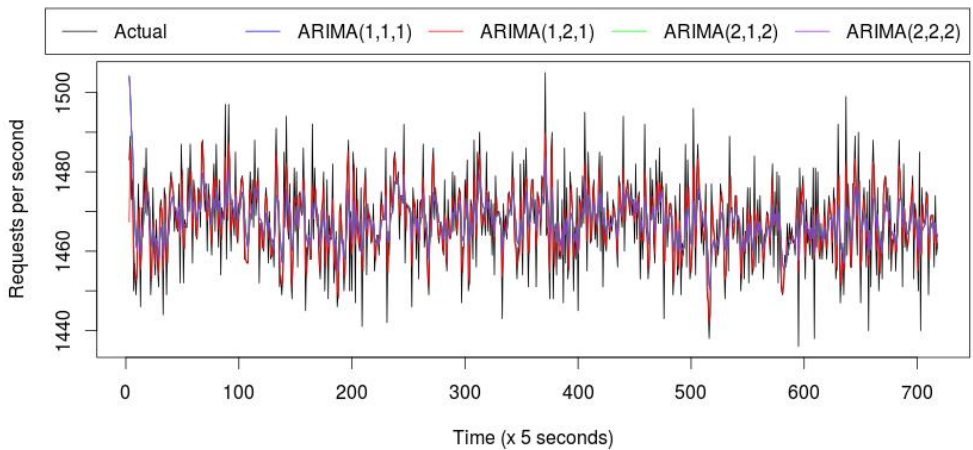

(b)

Fig. 2. Accuracy of different ARIMA models. (a): all the training period. (b): 1 hour snapshot.

TABLE I
PERFORMANCE OF DIFFERENT ARIMA MODELS.

| Model | MSE | Norm. VM hours | Norm. rejection |
|---|---|---|---|
| ARIMA (1,1,1) | 82.89775 | 0.7178 | 0.94 |
| ARIMA (1,2,1) | 55.75667 | 0.6518 | 0.92 |
| ARIMA (2,1,2) | 82.93857 | 0.6158 | 0.96 |
| ARIMA (2,2,2) | 82.80875 | 0.58953 | 1.00 |

Results are shown in Table I, which shows that a better fitting results in a better resource utilization: although the model with smallest MSE does not result in the minimum resource usage, it results in the minimum rejection rate, what means that the number of resources provisioned was the one that, among the alternatives, provided the best QoS: two models led to smaller number of resources provisioned, at a cost of higher rejection (because the available machines were not enough to handle all the requests) whereas one model resulted in higher VM utilization and slightly higher rejection rate, meaning that extra VMs were not provisioned in appropriate times.

## VIII. FUTURE DIRECTIONS

The complete realization of the potential of proposed architectural framework and goals require further investigation in the area of anomaly detection and prevention. The key challenges are outlined below.

- *Dealing with CPU spikes:* efficiently dealing with CPU spikes requires the availability (or development) of the resource consumption model of the application that can efficiently detect CPU load anomalies in a timely manner. Workload anomaly detection methods that use Markov chain models, although having several advantages, are not capable of dealing with CPU hogs as they are time consuming. One can investigate a regression-based transaction model to detect anomalies in a timely manner. Alternatively, one can look into Deep Learning approaches that have shown good potential in detecting anomalies in cloud environments. However, they have to be further improved to handle unseen anomalies such as One-Class SVMs.
- *Root cause analysis:* in real-world scenarios, changes in one application tier often can affect other tiers. Therefore, mining dependencies between anomalies of different application tiers is another promising research direction. Once obtained, they can be modeled and stored with the help of knowledge representation languages in the system. The knowledgebase can be later used to identify the root cause of anomalies or detect anomalies faster.
- *System metric anomaly detection versus workload anomaly detection (black, gray, or white box):* it would be interesting to compare the performance of systems with approaches that perform anomaly detections on workloads (request arrival time) or systems that consider resource consumption anomalies. Workload anomaly detection tends to provide an effective method when applied in web applications. The reason is that it enables prediction of how the load transfer from one node to another and therefore how an anomaly in load and resource consumption in one tier can lead to anomaly in the next tier.
- *Multi-resource anomaly detection:* considering multiple resources in anomaly detection has several advantages, as it is important to find out which resource contributes more to anomalies that are detected in application QoS. Considering only one resource at a time causes an unnecessary delay that can be prevented by checking distances among ranks of metrics and triggering scaling of CPU and memory simultaneously.

## IX. CONCLUSIONS

The sheer volume of structured and unstructured data generated by machines and humans give raise to the Big Data era. Businesses in many sectors such as finance, marketing, retailing, insurance, and real estate are just starting to leverage these data for commercial advantage. Similarly, governments and organizations are starting to build smart cities and e-health solutions that leverage Big Data to improve quality of life of the population. It is natural that the ICT industry—which supplies the underlying capability to enable Big Data—would also leverage it for its own benefit.

In this paper, we presented the challenges and opportunities of enhancing the operations of cloud data centers via Big Data analytics. Cloud data centers usually contains thousands to tens of thousands of physical (computing and networking hardware) and virtual (virtual machines and virtualized network functions) elements that are used by a variable number of users subject to SLAs. To enable services to comply with SLAs with minimum resource usage, techniques such as anomaly detection and prediction, regression and prediction of workloads and performance, and clustering can be used. Each of these techniques has been discussed, and an architectural framework for anomaly detection and prevention has been proposed.

Finally, a list of open issues and future research directions are identified. They show that there are still many open questions that need to be addressed to enable cloud infrastructures to get the most of Big Data analytics.